\documentclass[ALICE,manyauthors]{cernphprep}

\usepackage[comma,square,numbers,sort&compress]{natbib}
\usepackage{hyperref}
\usepackage{lineno}
\usepackage{xspace}
\usepackage{xcolor}
\usepackage{textcomp}
\usepackage[T1]{fontenc}
\usepackage{placeins}
\usepackage{multirow}

\newcommand{\snn}{\ensuremath{\sqrt{s_{\mathrm{NN}}}}\xspace}

\newcommand{\Hee}{\ensuremath{^{3}\mathrm{He}}\xspace}

\newcommand{\permille}{\textperthousand\xspace}

\newcommand{\pt}   {\ensuremath{p_{\mathrm{T}}}\xspace}
\newcommand{\mt}   {\ensuremath{m_{\mathrm{T}}}\xspace}

\newcommand{\dd}     {\ensuremath{\mathrm{d}}\xspace}
\newcommand{\dndy} {\mbox{\ensuremath{\dd N/\dd y}}\xspace}
\newcommand{\dndeta} {\mbox{\ensuremath{\dd N_\mathrm{ch}/\dd \eta}}\xspace}

\newcommand{\gmom}   {\mbox{\ensuremath{\mathrm{GeV}/c}}\xspace}
\newcommand{\dedx}{\mbox{\ensuremath{\dd E/\dd x}}\xspace}

\begin{document}%

\begin{titlepage}
\PHyear{2021}
\PHnumber{250}      
\PHdate{01 December}  
%

\title{Production of light (anti)nuclei in pp collisions at $\sqrt{s}~=~5.02$~TeV}
\ShortTitle{Production of light (anti)nuclei in pp collisions at  $\sqrt{s}~=~5.02$~TeV}   

\Collaboration{ALICE Collaboration\thanks{See Appendix~\ref{app:collab} for the list of collaboration members}}
\ShortAuthor{ALICE Collaboration} 

\begin{abstract}
The study of the production of nuclei and antinuclei in pp collisions has proven to be a powerful tool to investigate the formation mechanism of loosely bound states in high-energy hadronic collisions. In this paper, the production of protons, deuterons and \Hee and their charge conjugates at midrapidity is studied as a function of the charged-particle multiplicity in inelastic pp collisions at $\sqrt{s}=5.02$ TeV using the ALICE detector. Within the uncertainties, the yields of nuclei in pp collisions at $\sqrt{s}=5.02$ TeV are compatible with those in pp collisions at different energies and to those in p--Pb collisions when compared at similar multiplicities. The measurements are compared with the expectations of coalescence and Statistical Hadronisation Models. The results suggest a common formation mechanism behind the production of light nuclei in hadronic interactions and confirm that they do not depend on the collision energy but on the number of produced particles.

\end{abstract}
\end{titlepage}
\setcounter{page}{2}

\section{Introduction}
\label{sec:intro}

Light (anti)nuclei are abundantly produced in ultrarelativistic heavy-ion collisions ~\cite{Adam:2015vda,Adler:2001uy,Adler:2004uy} at the Large Hadron Collider (LHC), but their measurement in pp collisions is challenging due to their lower production yields. As a consequence, until few years ago there were only few measurements of the production rates of (anti)nuclei in small collision systems~\cite{Alper:1973my,Henning:1977mt,Adam:2015vda,Acharya:2017fvb}. This has recently changed thanks to the large pp data samples collected by ALICE at the LHC, which allow us to perform more precise and differential measurements of the production of light (anti)nuclei. In this paper, we present the detailed study of the multiplicity and transverse momentum dependence of (anti)proton, (anti)deuteron and (anti)$^3$He production in pp collisions at  $\sqrt{s} =$ 5.02 TeV. The results shown in the following are the most accurate obtained so far in small systems and represent the full compilation of data available for pp collisions at different energies at the end of the LHC Run 2.

The production mechanism of light (anti)nuclei in high-energy hadronic collisions is not fully understood. The classes of models used for comparison with the experimental results are the Statistical Hadronisation Models (SHM) and the coalescence models. SHMs assume that particles originated from an excited region evenly occupy all the available states in phase space~\cite{Hagedorn:346206}. Pb--Pb collisions, characterised by a large extension of the particle-emitting source and hence considered as large systems, are described according to a grand canonical ensemble~\cite{BRAUNMUNZINGER201676}. On the contrary, pp and p--Pb collisions, which are characterised by a small size and are considered as small systems, must be described based on a canonical ensemble, requiring the local conservation of the appropriate quantum numbers~\cite{Vovchenko:2019kes}. The expression Canonical Statistical Model (CSM) is used to underline the canonical description.

An important observable that provides information on the production mechanism is the ratio between the \pt-integrated yields of nuclei and protons. The measured d/p and $^3$He/p ratios show a rather constant behaviour as a function of centrality in Pb--Pb collisions. In contrast to that, they increase in pp and p--Pb collisions with increasing multiplicity, finally reaching the values measured in Pb--Pb collisions~\cite{Adam:2015vda,Acharya:2019rgc,Acharya:2019rys}. 
The constant nuclei-to-proton ratios in large collision systems is predicted by the SHMs~\cite{Sharma:2018jqf}, while the experimentally determined difference between small and large systems can be qualitatively explained as an effect of the canonical suppression of the nuclei yields for small system sizes. The prediction of the CSM saturates towards the grand canonical value at larger system size~\cite{Vovchenko:2018fiy} .

In coalescence models, (anti)nuclei are formed by nucleons close in phase space 
\cite{Kapusta:1980zz}.
In this approach, the coalescence parameter $B_{\mathrm{A}}$ relates the production of (anti)protons to the one of $\text{(anti)nuclei}$. $B_{\mathrm{A}}$ is defined as
\begin{equation}
    B_{\mathrm{A}}\left(p_{\mathrm{T}}^{\mathrm{p}}\right) = \frac{1}{2\pi p_{\mathrm{T}}^{\mathrm{A}}}\frac{\mathrm{d}^2N_{\mathrm{A}}}{\mathrm{d}y\mathrm{d}p^{\mathrm{A}}_{\mathrm{T}}} \; \bigg/ \left(\frac{1}{2\pi p_{\mathrm{T}}^{\mathrm{p}}}\frac{\mathrm{d}^2N_{\mathrm{p}}}{\mathrm{d}y\mathrm{d}\pt^{\mathrm{p}}}\right)^{\mathrm{A}} ,
\label{eq:BA}
\end{equation}
where \pt is the transverse momentum, $y$ the rapidity and $N$ the number of particles. The labels p and $A$ are used to denote properties related to protons and nuclei with mass number $A$, respectively. The production spectra of the $\text{(anti)protons}$ are evaluated at the transverse momentum of the nucleus divided by the mass number, so that $p_{\mathrm{T}}^{\mathrm{p}} = p_{\mathrm{T}}^{\mathrm{A}} /A$. Neutron spectra are assumed to be equal to proton spectra, due to the isospin symmetry restoration in hadron collisions at the LHC.
Since the coalescence process is expected to occur at the late stages of the collision, the $B_{\mathrm{A}}$ parameter is related to the emission volume. In a simple coalescence approach, which describes the uncorrelated particle emission from a point-like source, $B_\mathrm{A}$ is expected to be independent of \pt~and multiplicity. 
In this context, the measurements of the nuclei-to-proton ratios and of the $B_\mathrm{A}$ parameters in pp collisions at $\sqrt{s} =$~5.02~TeV reported in this paper are important to complete the present picture of the production of light nuclei in small systems. In addition, the increased statistics exploited in the present analysis will allow us to better constrain the models, thus to provide important inputs to both the theoretical and experimental communities.

\section{The ALICE apparatus}
\label{sec:detector}
A detailed description of the ALICE detectors can be found in~\cite{Abelev:2014ffa,ALICE:2008ngc} and references therein. In the following more information is given on the sub-detectors used to perform the analysis presented in this work, namely the V0, the Inner Tracking System (ITS), the Time Projection Chamber (TPC) and the Time-of-Flight (TOF). All of them are located inside a solenoidal magnet creating a magnetic field parallel to the beam line, with an intensity of 0.5~T for the data sample here considered. 

The V0 detector~\cite{Abbas:2013taa} is formed by two arrays of scintillation counters placed around the beam pipe on either side of the interaction point. They cover the pseudorapidity ranges $2.8 \leq \eta \leq 5.1$~\mbox{(V0A)}  and $-3.7 \leq \eta \leq -1.7$~\mbox{(V0C)}.
The collision multiplicity is estimated using the signal amplitude in the V0 detector, which is also used as a trigger detector. More details will be given in Section~\ref{sec:eventselection}.
 
The ITS~\cite{Aamodt:2010aa}  provides high resolution track points in the proximity of the interaction region and consists of three subsystems. Going from the innermost to the outermost subsystem, we find: two layers of Silicon Pixel Detectors (SPD), two layers of Silicon Drift Detectors (SDD) and two layers  equipped with double-sided Silicon Strip Detectors (SSD). The ITS extends radially from 3.9 cm to 43 cm, it is hermetic in azimuth and it covers the pseudorapidity range $|\eta |<0.9$.

The same pseudorapidity range is covered by the TPC~\cite{Alme:2010ke}, which is the main tracking detector, consisting of a hollow cylinder whose axis coincides with the nominal beam axis. The active volume, filled with a Ne/CO$_2$/N$_2$ gas mixture at atmospheric pressure, has an inner radius of about 85~cm and an outer radius of about 250~cm.
The trajectory of a charged particle is estimated using up to 159 combined measurements (clusters) of drift times and radial positions of the ionisation electrons.
The charged-particle tracks are then reconstructed by combining the hits in the ITS and the measured clusters in the TPC. 
The TPC is also used for particle identification (PID) by measuring the specific energy loss (\dedx) in the TPC gas. In pp collisions, the \dedx in the TPC is measured with a resolution of $\approx 5.2\%$~\cite{Abelev:2014ffa}.

The TOF~\cite{Akindinov:2013tea} covers the full azimuth for the pseudorapidity interval $|\eta|<0.9$. The detector is based on the Multigap Resistive Plate Chambers (MRPC) technology and is located, with a cylindrical symmetry, at an average distance of 380 cm from the beam axis.
The particle identification is based on the difference between the measured time of flight and its expected value, computed for each mass hypothesis from track momentum and length. A precise starting signal for the measurement of the time of flight by the TOF is provided by the T0 detector, consisting of two arrays of Cherenkov counters, T0A and T0C, which cover the pseudorapidity regions $4.61 \leq \eta \leq 4.92$ and $3.28 \leq \eta \leq 2.97$, respectively~\cite{Adam:2016ilk}. The overall resolution on the particles time of flight, including the start time, is $\approx$~80~ps.

\section{Data sample}
\label{sec:eventselection}
This analysis is based on approximately 900 million pp collisions (events) at $\sqrt{s}=5.02$ TeV collected in 2017 by ALICE at the LHC. Events are selected by a minimum-bias (MB) trigger, requiring at least one hit in each of the two V0 detectors. An additional offline rejection is performed to remove events with more than one reconstructed primary vertex (pile-up events) and events triggered by interactions of the beam with the residual gas in the LHC beam pipe~\cite{Abbas:2013taa}. In total, 1.8\% of the collected events are rejected due to these selections.

The production of (anti)nuclei is measured around midrapidity, within a rapidity range of \mbox{$|y|<0.5$}, and within the pseudorapidity interval $|\eta|<0.8$ to maximise the detector performance. The selected tracks are required to have at least 70 reconstructed points in the TPC and two points in the ITS in order to guarantee good track momentum and \dedx resolution in the relevant \pt ranges. In addition, at least one hit in the SPD is required to ensure a resolution of the distance of closest approach to the primary vertex better than 300 $\mu$m, both along the beam axis (DCA$_\mathrm{z}$) and in the transverse plane (DCA$_\mathrm{xy}$)~\cite{Abelev:2014ffa}. The quality of the accepted tracks is checked by requiring the $\chi^2$ per TPC reconstructed point and per ITS reconstructed point to be less than 4 and 36, respectively. Finally, tracks originating from kink topologies of kaon and pion decays are rejected.

Data are divided into multiplicity intervals classified by a roman numeral from I to X, going from the highest to the lowest multiplicity~\cite{Acharya:2019rgc}. In order to achieve a higher statistical precision, classes are merged into nine classes for (anti)protons and (anti)deuterons and into two classes for (anti)helion. The multiplicity classes are defined from the mean of the V0 signal amplitudes as percentiles of the $\mathrm{INEL}>0$ pp cross section, where $\mathrm{INEL}>0$ events are defined as collisions with at least one charged particle in the pseudorapidity region $|\eta|<1$~\cite{Acharya:2019kyh}. The mean charged-particle multiplicities for each class, $\left<\dndeta\right>$, are listed in Table~\ref{tab:yields}.

\begin{table}
\centering
\caption{Multiplicity classes for the different measurements, with the corresponding charged-particle multiplicity density at midrapidity $\langle$d$N_\mathrm{ch}$/d$\eta\rangle$ and percentiles of the INEL $>$ 0 pp cross section, and \pt-integrated yields d$N$/d$y$ for the different species. For protons, statistical uncertainties are negligible with respect to systematic uncertainties.}
\label{tab:yields}
\begin{tabular}{cccccc}
\hline
\multirow{2}{*}{Class} & V0 & \multirow{2}{*}{$\langle$d$N_\mathrm{ch}$/d$\eta\rangle_{|\eta_{\mathrm{lab}}|<0.5}$} & \multicolumn{3}{c}{d$N$/d$y$}                                                           \\
                       & percentile    &                                                & p ($\times~10^{-1}$)                         & d ($\times~10^{-4}$)       & ${}^{3}$He ($\times~10^{-7}$) \\ \hline
I & 0 -- 1\%                    & 18.5 $\pm$ 0.2                                     & 5.0 $\pm$ 0.0 $\pm$ 0.3 & 10.7 $\pm$ 0.2 $\pm$ 0.7   &                               \\
II & 1 -- 5\%                     & 14.5 $\pm$ 0.2                                     & 4.0 $\pm$ 0.0 $\pm$ 0.2 & 8.10 $\pm$ 0.07 $\pm$ 0.39 &                               \\
III & 5 -- 10\%                   & 11.9 $\pm$ 0.2                                     & 3.4 $\pm$ 0.0 $\pm$ 0.2 & 6.36 $\pm$ 0.05 $\pm$ 0.32 &                               \\
IV -- V & 10 -- 20\%                 & 9.7 $\pm$ 0.1                                      & 2.8 $\pm$ 0.0 $\pm$ 0.2 & 4.92 $\pm$ 0.03 $\pm$ 0.24 &                               \\
VI & 20 -- 30\%                    & 7.8 $\pm$ 0.1                                      & 2.2 $\pm$ 0.0 $\pm$ 0.1 & 3.60 $\pm$ 0.03 $\pm$ 0.18 &                               \\
VII & 30 -- 40\%                    & 6.3 $\pm$ 0.1                                      & 1.8 $\pm$ 0.0 $\pm$ 0.1 & 2.65 $\pm$ 0.03 $\pm$ 0.14 &                               \\
VIII & 40 -- 50\%                  & 5.2 $\pm$ 0.1                                      & 1.5 $\pm$ 0.0 $\pm$ 0.1 & 1.98 $\pm$ 0.02 $\pm$ 0.09 &                               \\
IX  & 50 -- 70\%                   & 3.9 $\pm$ 0.1                                      & 1.1 $\pm$ 0.0 $\pm$ 0.1 & 1.28 $\pm$ 0.01 $\pm$ 0.06 &                               \\
X  & 70 -- 100\%                      & 2.4 $\pm$ 0.1                                      & 0.6 $\pm$ 0.0 $\pm$ 0.1 & 0.48 $\pm$ 0.01 $\pm$ 0.06 &                               \\ \hline
I -- III    & 0 -- 10\%              & 13.6 $\pm$ 0.2                                     &                            &                            & 5.4 $\pm$ 0.3 $\pm$ 0.7       \\
IV -- X   & 10 -- 100\%                & 4.9 $\pm$ 0.1                                      &                            &                            & 1.5 $\pm$ 0.1 $\pm$ 0.4       \\ \hline
INEL > 0 & 0 -- 100\%                    & 5.5 $\pm$ 0.1                                      & 1.5 $\pm$ 0.0 $\pm$ 0.1 & 2.29 $\pm$ 0.01 $\pm$ 0.12 & 1.7 $\pm$ 0.1 $\pm$ 0.4      
\end{tabular}
\end{table}

\section{Data analysis}
\label{sec:analysis}

\subsection{Raw yield extraction}
The first important step in the analysis is the particle identification. As already shown in previous works~\cite{Adam:2015vda,Acharya:2017fvb,Acharya:2019rgc,Acharya:2020sfy,ALICE:2021mfm}, the identification of (anti)nuclei is performed with two different methods, depending on the particle species and on the transverse momentum. For (anti)protons and (anti)deuterons with \pt < 1 GeV/\textit{c}, the identification relies on the measurement of the \dedx using the TPC. The number of signal candidates is extracted through a fit with a Gaussian with two exponential tails to the $n_{\sigma_{\mathrm{TPC}}}$ distribution for each \pt interval. The $n_{\sigma_{\mathrm{TPC}}}$ is defined as the difference between the measured and the expected \dedx for each particle species, divided by \dedx resolution of the TPC. For \pt $\ge$ 1 GeV/$c$, it is more difficult to separate (anti)protons and (anti)deuterons from other charged particles of $|Z|=1$. Therefore, PID is performed using the TOF detector information in addition. The squared mass of the particle is evaluated as $m^{2} = p^{2}\left(t_{\mathrm{TOF}}^2/L^2 - 1/c^2\right)$, where $t_\mathrm{TOF}$ is the measured time of flight, $L$ is the length of the track and $p$ is the momentum of the particle. In order to reduce the background, the tracks are in addition required to have $|n_{\sigma_{\mathrm{TPC}}}| < 3$. The squared mass distributions of the signal are fitted with a Gaussian function with an exponential tail. Background originating from other particle species or from the random match of a TOF hit with another track significantly increases with \pt and is modelled with the sum of Gaussian and exponential functions. For (anti)helion, only the TPC \dedx measurement is used, because their signal in the TPC can be easily separated from the one of other particle species, due to the electric charge (Z = 2). The raw yield of (anti)helion is obtained through a fit of the $n_{\sigma_{\mathrm{TPC}}}$ with a Gaussian function for the signal and a Gaussian function for the contamination coming from (anti)triton, where present. When the background is negligible, the raw yield is extracted by directly counting the (anti)nuclei candidates. Otherwise, the TPC \dedx and TOF squared mass distributions are fitted with the aforementioned models, using an extended-maximum-likelihood approach and the yield is obtained as a fit parameter. In the signal extraction, the fit quality is monitored and a successful Pearson test is required with the probability to reject a true hypothesis of $5\%$.
\subsection{Efficiency and acceptance correction}
\label{sec:efficiency}
The raw yield must be corrected to take into account the tracking efficiency and the detector acceptance. This correction is evaluated from Monte Carlo (MC) simulated events, which are generated using the event generator PYTHIA8.21 (Monash2013 tune)~\cite{sjostrand2008brief}. However, since PYTHIA8 does not handle the production of nuclei properly, it is necessary to inject (anti)nuclei on top of each generated event. In each pp collision, one deuteron, one antideuteron, one helion or one antihelion are injected, randomly chosen from a flat rapidity distribution in the range $|y|<1$ and a flat \pt distribution in the range \pt$\in [0,10] $ GeV/$c$. The GEANT4~\cite{Geant4} transport code is exploited to describe the hadronic interaction of the particles propagating through the detector material. The correction is defined as the ratio between the number of reconstructed (anti)nuclei in the rapidity range $|y|<0.5$ and in the pseudorapidity interval $|\eta|<0.8$ and the number of generated ones in $|y|<0.5$. The correction is computed separately for each (anti)nucleus and for the TPC and TOF analyses. Moreover, the raw signal needs to be corrected for trigger inefficiencies. The selected events are requested to have at least one charged-particle in the pseudorapidity region $|\eta|<1$ (INEL $>$ 0)~\cite{Acharya:2019kyh}. Some INEL $>$ 0 events can be lost due to the finite trigger efficiency (event loss) and all the particles produced in those events are lost as well (signal loss). Hence, it is necessary to correct the spectra for the event and the signal losses. The correction must be evaluated from MC simulations because the number of rejected events and lost particles is only known there. For (anti)protons, this correction is directly computed from the MC simulation because their production is handled by the event generator. On the contrary, (anti)nuclei are injected on top of a pp collision and a direct estimation from the MC is not possible, because there would be a bias in the number of lost (anti)nuclei. For this reason, the correction for pions, kaons and protons is evaluated in this case in a different MC data set with no injected nuclei and the average value is used for (anti)deuterons and (anti)helions. Further details on this method can be found in~\cite{Acharya:2019rgc, Acharya:2020sfy}. This correction is negligible at high multiplicity ($<$~1\permille) and becomes relevant at low multiplicity (up to 14\% for (anti)protons and (anti)deuterons, 2\% for (anti)helions, in the low \pt region $p_\mathrm{T}<1$ GeV/$c$).

\subsection{Secondary (anti)nuclei contamination}
The contribution of secondary (anti)nuclei, i.e. (anti)nuclei that are not produced directly in the collision, must be subtracted from the total measured yields. Secondary nuclei are mostly produced in the interaction of particles with the vacuum beam pipe and the detector material. Moreover, an important contribution to secondary (anti)protons is also given by the weak decay of heavier particles. All particles coming from strong and electromagnetic decays are considered as primary. (Anti)deuterons and (anti)helions receive a negligible background contribution from weak decays, since the only known contribution comes from the decays of hypertriton ($^3_\Lambda$H $\rightarrow$ d + p + $\pi$ and $^3_\Lambda$H $\rightarrow$ $^3$He + $\pi$) and their antimatter counterparts, whose production is known to be suppressed in pp collisions~\cite{Acharya:2017fvb}. Finally, the production of secondary antideuterons and antihelions from material is extremely rare due to baryon number conservation. The fraction of primary (anti)nuclei is evaluated through a template fit to the DCA$_\mathrm{xy}$ distribution of the data, as described in~\cite{Adam:2015vda}. The templates for primary and secondary (anti)protons and deuterons are obtained from MC simulations. For (anti)protons, two templates are used to describe both (anti)protons from weak decays and from material. While the template for primary (anti)helions is extracted from the MC as well,  this is not possible for the template for secondaries, due to the very rare production of antihelion. For this reason, the (anti)proton template at half the (anti)helion \pt is used as a proxy for the (anti)helion one. This procedure is based on the assumption that the DCA$_\mathrm{xy}$ distributions of secondary (anti)helions can be represented by the DCA$_\mathrm{xy}$ distributions of (anti)protons at a transverse momentum which is scaled with the rigidity $p/z$ of (anti)helion, where $z$ is the (anti)helion electric charge. The contribution of secondary nuclei is observed to be more relevant at low \pt (20\% for protons, 40\% for deuterons and 90\% for helions) and to decrease exponentially with increasing transverse momentum.

\subsection{Systematic uncertainties}
One contribution of the systematic uncertainties comes from the adopted track selection criteria. This uncertainty is evaluated by varying the selections, as done in~\cite{Acharya:2019rgc}. The effect of the subtraction of secondary (anti)nuclei is studied with the variation of the DCA$_\mathrm{z}$ and DCA$_\mathrm{xy}$ selections as well. This is the most relevant contribution for (anti)helion at low \pt, decreasing with \pt. The estimation of the systematic uncertainty related to the raw signal extraction depends on the considered species. For (anti)protons, the difference between the signal extracted by direct count and the one extracted from the fit is taken into account. For (anti)deuterons, this is obtained by varying the interval in which the direct counting of (anti)deuterons is performed. Finally, for (anti)helion a toy MC has been developed in order to generate 10000 TPC \dedx samples that are compatible with the default one. A possible bias in the signal extraction process is investigated by refitting each distribution and looking into the variation of the extracted yields. Another source of systematic uncertainty is given by the incomplete knowledge of the material budget of the detector in the MC simulations. This is evaluated by comparing different MC simulations in which the material budget of the ALICE detector was varied by $\pm$4.5\%~\cite{Abelev:2014ffa} after conversions. This value corresponds to the uncertainty on the determination of the material budget obtained by measuring photon conversions. The imperfect knowledge of the hadronic interaction cross section of (anti)nuclei in the material contributes to the systematic uncertainty as well and depends on the particle species. Similarly, an uncertainty related to the ITS-TPC matching is considered and evaluated from the difference between the ITS-TPC matching efficiencies in data and MC. Finally, the trigger inefficiency is also a source of systematic uncertainties. The uncertainty is assumed to be half of the difference between the signal loss correction (described in section~\ref{sec:efficiency}) and unity. It strongly depends on the event multiplicity: it is negligible at high multiplicity and contributes up to 7\% in the lowest event class for (anti)deuterons and (anti)helions. Where present, it decreases with increasing \pt. The list of all the sources of systematic uncertainty for the INEL $>$ 0 multiplicity class is reported in Table~\ref{tab:syst}. The average values between matter and antimatter are reported for (anti)protons, (anti)deuterons and (anti)helions, for the lowest and highest \pt values of the measured spectra.

\begin{table}[h]
\centering
\caption{Summary of the contributions to the systematic uncertainties of the yield for the INEL $>$ 0 event class for the different species.}
\label{tab:syst}
\renewcommand{\arraystretch}{1.2}
\begin{tabular}{lcccccc}
\hline
                   & \multicolumn{2}{c}{p} & \multicolumn{2}{c}{d} & \multicolumn{2}{c}{$^3$He} \\
$p_\mathrm{T}$ (GeV/$c$) & 0.3       & 3.5       & 0.7       & 3.4       & 0.9          & 4.2         \\ \hline
Track selection          & < 1\%     & 9.5\%   & < 1\%    & 2\%     & < 1\%       & 4\%       \\
Secondary particles         & 3.5\%   & 5\%     & 1\%     & < 1\%    & 16\%       & 2.5\%     \\
Signal extraction        & 1\%     & 1.5\%   & < 1\%    & 7.5\%   & < 1\%       & 4\%       \\
Material budget          & 2\%     & < 1\%    & < 1\%    & < 1\%    & 4.5\%      & < 1\%      \\
Hadronic interaction     & < 1\%   & < 1\%   & 1.5\%    & 2\%     & 1\%        & < 1\%       \\
ITS-TPC matching         & 1\%     & 2.5\%   & 1\%     & 2.5\%   & 2\%        & 2.5\%     \\
Trigger inefficiency              & 2\%     & < 1\%    & < 1\%    & < 1\%    & < 1\%       & < 1\%      \\ \hline
Total                    & 4.5\%   & 11\%    & 3\%     & 9\%   & 17\%     & 7\%       \\ \hline
\end{tabular}
\end{table}

\FloatBarrier

\section{Results and discussion}
\label{sec:results}

\begin{figure}[ht]
	\centering
	\includegraphics[width=\textwidth]{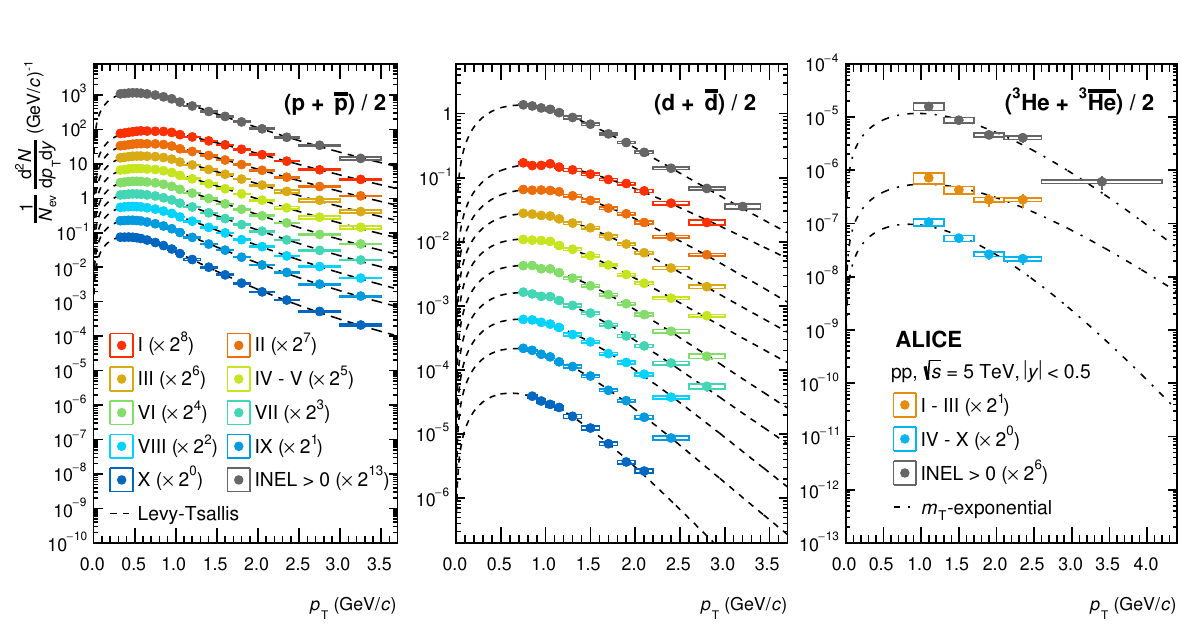}
     \caption{Transverse-momentum spectra of (anti)protons (left), (anti)deuterons (center) and (anti)helions (right) in the different multiplicity classes, reported in Table~\ref{tab:yields}. (Anti)deuteron and (anti)proton spectra are fitted with a L\'evy-Tsallis function~\cite{Tsallis:1987eu}, while (anti)helion spectra are fitted with an exponential function with respect to the transverse mass \mt.}
     \label{fig:spectra}
\end{figure}

\begin{figure}[ht]
	\centering
	\includegraphics[width=\textwidth]{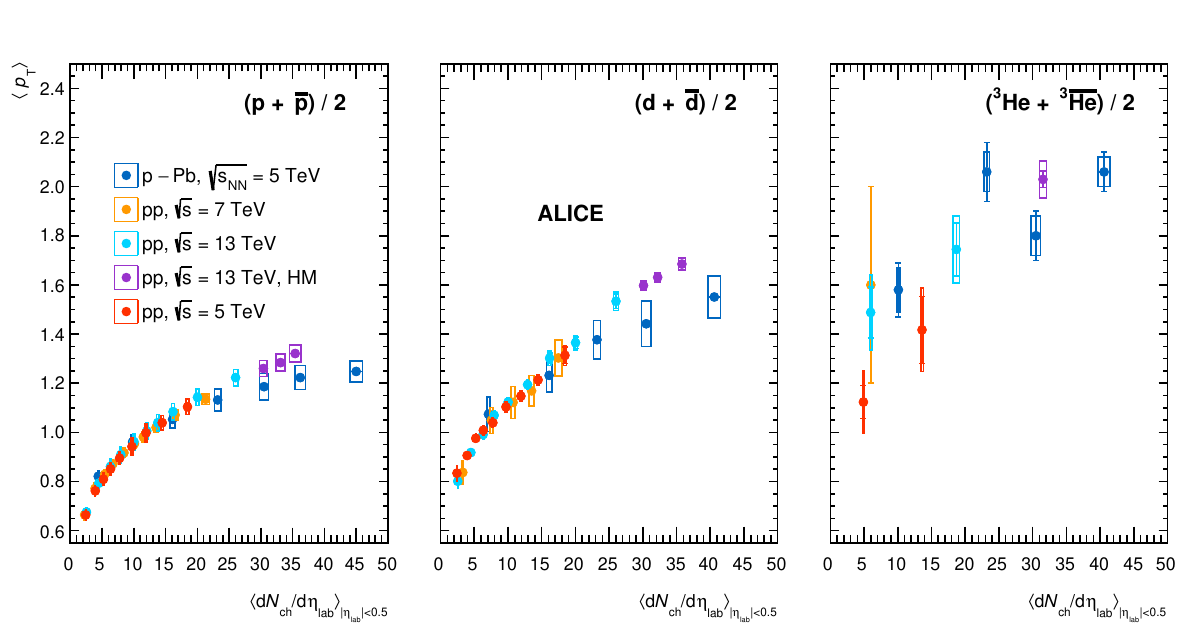}
     \caption{Mean transverse momentum of (anti)protons (left), (anti)deuterons (centre) and (anti)helions (right) in pp collisions at $\sqrt{s}~=~5.02$~TeV, in high-multiplicity pp collisions at $\sqrt{s}~=~13$~TeV~\cite{ALICE:2021mfm}, in INEL $>$ 0 pp collisions at $\sqrt{s}~=~13$~TeV~\cite{ALICE:2020nkc,Acharya:2020sfy,ALICE:2021mfm} and at $\sqrt{s}~=~7$~TeV~\cite{ALICE:2018pal,Acharya:2019rgc,Acharya:2017fvb}, and in p--Pb collisions at $\snn~=~5.02$~TeV~\cite{ALICE:2013wgn,Acharya:2019rys,3He_pPb}. The statistical uncertainties are represented by vertical bars while the systematic uncertainties are represented by boxes.}
     \label{fig:meanpt}
\end{figure}

The transverse-momentum spectra for (anti)protons, (anti)deuterons and (anti)helions are shown in Fig.~\ref{fig:spectra}. In each \pt interval, the reported yield is the average between matter and antimatter. Both of them are compatible, as already observed in previous measurements carried out by ALICE~\cite{Acharya:2020sfy,Acharya:2019rgc,Acharya:2019rys,Adam:2015vda}. The measured spectra are fitted in order to extrapolate the yields in the unmeasured \pt-region. For (anti)protons and (anti)deuterons, data are fitted with a L\'evy-Tsallis function~\cite{Tsallis:1987eu}, while for (anti)helion a simple exponential depending on \mt is used because it provides a better description of the data. The fraction of the yield obtained from the extrapolation depends on the considered particle species and on the multiplicity class, since the \pt-coverage is generally different, being maximum (minimum) at high (low) multiplicity. For (anti)protons, the extrapolation contributes with a fraction of 10\% (20\%) of the total yield for the highest (lowest) multiplicity class, while for (anti)deuterons and (anti)helions it contributes with a fraction of 25\% (55\%) and 35\% (40\%) of the total yield, respectively. The \pt-spectra are also fitted with a Boltzmann function and a simple exponential depending on \pt, in order to quantify the effect of the chosen function on the \pt-integrated yield. The difference between the yields obtained with the reference and the alternative functions is taken as systematic uncertainty. This accounts for $\approx$ 2\% for (anti)protons and (anti)deuterons, depending on the transverse-momentum coverage of the spectra, whereas for (anti)helions this accounts for 12\% in the highest multiplicity class and $\approx$ 19\% in the lowest multiplicity class. The \pt-integrated yields \dndy are reported in Table~\ref{tab:yields}. For (anti)protons, the statistical uncertainties on the yields are negligible, being $\approx$1\% of the systematic uncertainty.  Figure~\ref{fig:meanpt} shows the mean transverse momentum $\langle\pt\rangle$ as a function of charged-particle multiplicity. The results are compared with those obtained in previous measurements and they confirm the increasing trend with multiplicity. Moreover, a clear mass ordering is present, as already observed for other light-flavoured particle species and for different collision systems and energies~\cite{ALICE:2013wgn,ALICE:2019avo}.

\begin{figure}[ht]
    \centering
	\includegraphics[width=\textwidth]{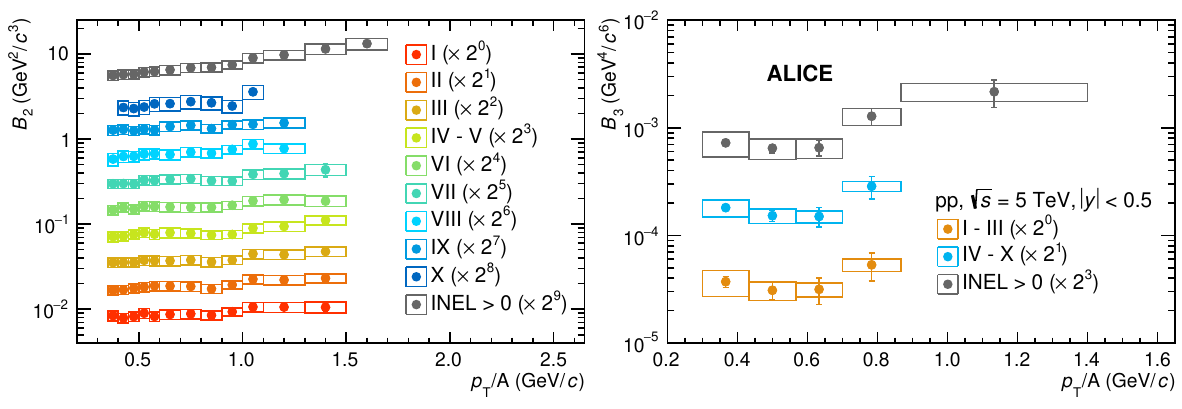}
     \caption{Coalescence parameters $B_2$ for (anti)deuterons (left) and $B_3$ for (anti)helions (right) for different multiplicity classes. The multiplicity decreases moving from the bottom up. The statistical uncertainties are represented by vertical bars while the systematic uncertainties are represented by boxes. $B_{\mathrm{A}}$ is shown as a function of $\pt/A$, being $A~=~2$ the mass number of deuteron and  $A~=~3$ the mass number of helion.}
     \label{fig:bA_pt}
\end{figure}

Combining the information from the production spectra of protons and nuclei, the coalescence parameter can be evaluated according to Eq.~\ref{eq:BA}. Figure~\ref{fig:bA_pt} shows the coalescence parameter as a function of transverse momentum for (anti)deuterons ($B_2$) and (anti)helions ($B_3$).
The $B_2$ and $B_3$ values in the fine multiplicity classes are consistent with a flat trend, while for the multiplicity-integrated sample the coalescence parameter increases with \pt. This behaviour was already observed in other measurements by ALICE in pp collisions~\cite{Acharya:2020sfy,Acharya:2019rgc} at different energies.
In particular, it is now understood that the increase with transverse momentum of the coalescence parameter in INEL $>$ 0 collisions is, in large part, due to the change in shape of the transverse momentum spectra of protons in different multiplicity intervals~\cite{Acharya:2019rgc}.
It is also worth mentioning that in pp collisions at high multiplicity (HM)~\cite{ALICE:2021mfm}, where the system size is larger than the one resulting from INEL > 0 collisions, the raise with \pt cannot be neglected even in fine multiplicity classes. In~\cite{ALICE:2021mfm}, it was shown that the $B_{\mathrm{A}}$ as a function of transverse momentum can be described by coalescence predictions, assuming a Gaussian wave function for the nuclei.

\begin{figure}[ht]
	\centering
    \includegraphics[width=\textwidth]{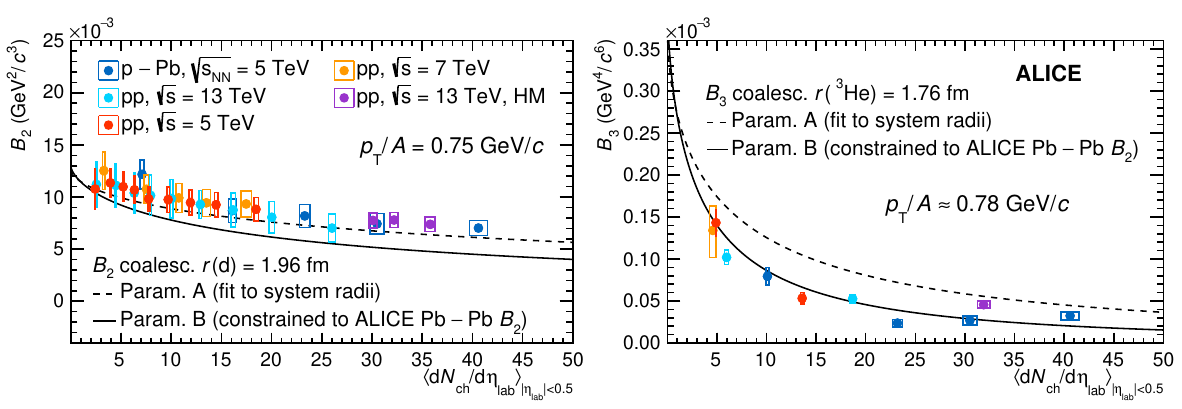}
     \caption{Left: $B_2$ as a function of multiplicity in INEL $>$ 0 pp collisions at $\sqrt{s}~=~5.02$~TeV, in high-multiplicity pp collisions at $\sqrt{s}~=~13$~TeV~\cite{ALICE:2021mfm}, in INEL $>$ 0 pp collisions at $\sqrt{s}~=~13$~TeV~\cite{Acharya:2020sfy} and at $\sqrt{s}~=~7$~TeV~\cite{Acharya:2019rgc}, and in p--Pb collisions at $\snn~=~5.02$~TeV~\cite{Acharya:2019rys}. Right: $B_3$ as a function of multiplicity in INEL $>$ 0 pp collisions at $\sqrt{s}~=~5.02$~TeV, in high-multiplicity pp collisions at $\sqrt{s}~=~13$~TeV~\cite{ALICE:2021mfm}, in INEL $>$ 0 pp collisions at $\sqrt{s}~=~13$~TeV~\cite{ALICE:2021mfm} and at $\sqrt{s}~=~7$~TeV~\cite{Acharya:2017fvb}, and in p--Pb collisions at $\snn~=~5.02$~TeV~\cite{3He_pPb}. The statistical uncertainties are represented by vertical bars while the systematic uncertainties are represented by boxes. The two lines are theoretical predictions of the coalescence model based on two different parameterisations of the system radius as a function of multiplicity.}
     \label{fig:BA_mult}
\end{figure}

Insights into the dependence of the production mechanisms on the system size can also be obtained by studying the evolution of $B_\mathrm{A}$ with charged-particle multiplicity. Indeed, as shown in~\cite{Bellini:2018epz}, the charged-particle multiplicity $\langle \mathrm{d}N_\mathrm{ch}/\mathrm{d}\eta \rangle$ can be considered as a proxy of the system size. Figure~\ref{fig:BA_mult} shows $B_2$ and $B_3$ as a function of charged-particle multiplicity for different collision systems and energies. The presented measurements are obtained in transverse momentum ranges with central values of $\pt/A=0.75~\gmom$ for $B_2$ and $\pt/A=0.78$ \gmom for $B_3$, but the trend is alike for other values.

\begin{figure}[ht]
	\includegraphics[width=\textwidth]{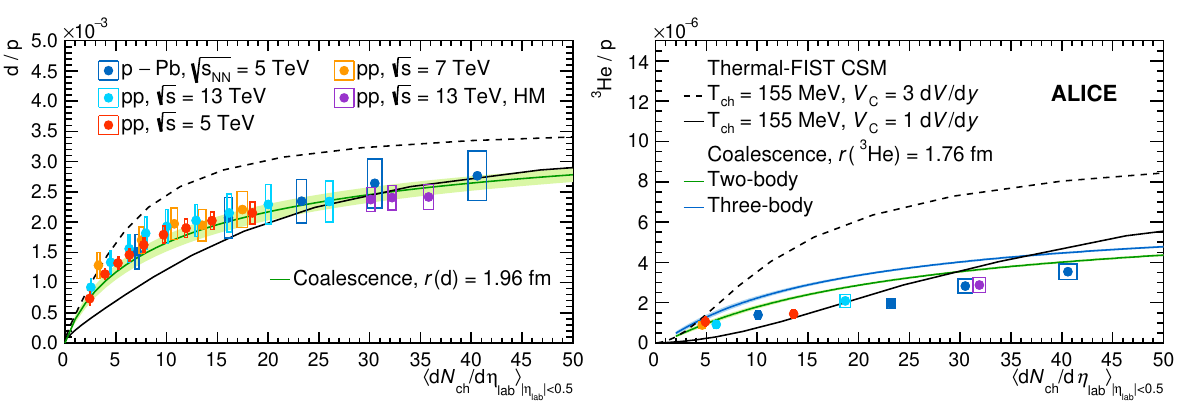}
     \caption{
     Ratio between the \pt-integrated yields of nuclei and protons as a function of multiplicity for (anti)deuterons (left) and (anti)helions (right). Measurements are performed in INEL $>$ 0 pp collisions at $\sqrt{s}~=~5.02$~TeV, in high-multiplicity pp collisions at $\sqrt{s}~=~13$~TeV~\cite{ALICE:2021mfm}, in INEL $>$ 0 pp collisions at $\sqrt{s}~=~13$~TeV~\cite{Acharya:2020sfy,ALICE:2021mfm} and at $\sqrt{s}~=~7$~TeV~\cite{Acharya:2017fvb}, and in p--Pb collisions at $\snn~=~5.02$~TeV~\cite{Acharya:2019rys,3He_pPb}. The statistical uncertainties are represented by vertical bars while the systematic uncertainties are represented by boxes. The two black lines are the theoretical predictions of the Thermal-FIST statistical model~\cite{Vovchenko:2018fiy} for two sizes of the correlation volume $V_\mathrm{C}$. For (anti)deuterons, the green band represents the expectation from a coalescence model~\cite{Sun:2018mqq}. For (anti)helion, the green and blue lines represent the expectations from a two-body and three-body coalescence models~\cite{Sun:2018mqq}.
     }
     \label{fig:AoP_mult}
\end{figure}
The measurements are compared with the theoretical predictions from~\cite{Bellini:2018epz}, where two different parameterisations of the source radius as a function of multiplicity are used (see~\cite{Bellini:2018epz} for details). It is evident that there is no single parameterisation of the system size that is able to fit both the measured $B_2$ and $B_3$. However, as stated also in~\cite{ALICE:2021mfm}, charged-particle multiplicity is not a perfect proxy for the system size, because for each multiplicity the source radius depends also on the transverse-momentum of the particle of interest. Anyhow, the data corresponding to the different collision systems and energies confirm a trend with multiplicity, which can be interpreted as an effect of the interplay between the size of the system and that of the nucleus.
Indeed, at low charged-particle multiplicity, the system size is comparable with the size of the nucleus (about $2~\mathrm{fm}$, depending on the nuclear species and on the parameterisation of the model), determining the slow decrease with multiplicity. On the contrary, increasing the multiplicity the system size becomes larger and larger than the nucleus size, making the coalescence process less and less probable~\cite{Adam:2015vda,Bellini:2018epz}.

Figure~\ref{fig:AoP_mult} shows the ratios between the \pt-integrated yields of nuclei and protons as a function of charged-particle multiplicity. A common trend as a function of the charged-particle multiplicity is seen, monotonically increasing for pp and p--Pb collisions and eventually saturating for Pb--Pb collisions~\cite{ALICE:2021mfm}. This is the effect of the interplay between the different evolution with the charged-particle multiplicity of the source size and of the particle yields~\cite{ALICE:2021mfm}. The systematic uncertainties in this analysis are reduced with respect to the previous ALICE measurements thanks to the recent studies on the interaction cross section of antideuteron with the material~\cite{ALICE:2020zhb}. The experimental data are compared with the predictions of both Thermal-FIST~\cite{Vovchenko:2018fiy} CSM and coalescence model~\cite{Sun:2018mqq}. The CSM prediction is provided for different correlation volumes $V_\mathrm{C}$, from 1 to 3 times the volume d$V$/d$y$. For both (anti)deuterons and (anti)helions, the CSM and the coalescence model can qualitatively describe the observed trend. A detailed study of the $V_\mathrm{C}$ value is required to determine if the CSM is able to describe simultaneously the deuteron and helion measurement here reported. The coalescence model seems to describe better the data points, and better for (anti)deuterons than for (anti)helions, where some tension at intermediate multiplicity is visible.

\section{Conclusions}
\label{sec:conclusions}
The LHC demonstrated to be an unprecedented antimatter factory. The production of nuclei and antinuclei has been explored at all energies delivered by the LHC during its Run 2~\cite{ALICE:2021mfm,3He_pPb,Acharya:2017fvb,Acharya:2019rys,Acharya:2019rgc,Acharya:2020sfy} and a clear pattern emerged: the production of nuclei is tightly driven by the underlying event multiplicity. Other variables, like the collision energy or even the colliding system (pp or p--Pb), are essentially irrelevant in the description of the nucleosynthesis processes in hadronic collision.

The CSM can explain qualitatively the observed trend in the nucleus-to-proton ratios as a function of multiplicity.
On the other hand, coalescence connects the hadron-emitting source size with the observed production of nuclei. The size of the hadron-emitting source increases with multiplicity and decreases with momentum as demonstrated by recent particle correlation measurements~\cite{ALICE:2020ibs}. Through this observation, coalescence can predict the  yield of nuclei as a function of both multiplicity and momentum starting from the measured proton spectrum.
In this paper, it is shown that the coalescence prediction agrees quantitatively with the measured deuteron-to-proton ratio, while the helion-to-proton ratio in pp collisions at 5.02 TeV confirms the trend of the previous measurements deviating from the coalescence prediction at intermediate multiplicities.
However, the comparison between the coalescence parameters with coalescence predictions show great sensitivity to different source size parameterisations, suggesting that some of the observed discrepancies might be due to the source size determination.
During the LHC Run 3, the ALICE experiment targets an integrated luminosity of 6 pb$^{-1}$ for pp collisions at 5.02 (or 5.5) TeV and up to 200 pb$^{-1}$ at 13 TeV~\cite{ALICE:2020fuk}, which corresponds to a sample larger by at least a factor 400 with respect to Run 2. This sample will enable a simultaneous study of the production of nuclei and the size of the system, similarly to what has already been done in high-multiplicity pp collisions at $\sqrt{s}=$ 13 TeV~\cite{ALICE:2021mfm}.

\newenvironment{acknowledgement}{\relax}{\relax}
\begin{acknowledgement}
\section*{Acknowledgements}

The ALICE Collaboration would like to thank all its engineers and technicians for their invaluable contributions to the construction of the experiment and the CERN accelerator teams for the outstanding performance of the LHC complex.
The ALICE Collaboration gratefully acknowledges the resources and support provided by all Grid centres and the Worldwide LHC Computing Grid (WLCG) collaboration.
The ALICE Collaboration acknowledges the following funding agencies for their support in building and running the ALICE detector:
A. I. Alikhanyan National Science Laboratory (Yerevan Physics Institute) Foundation (ANSL), State Committee of Science and World Federation of Scientists (WFS), Armenia;
Austrian Academy of Sciences, Austrian Science Fund (FWF): [M 2467-N36] and Nationalstiftung f\"{u}r Forschung, Technologie und Entwicklung, Austria;
Ministry of Communications and High Technologies, National Nuclear Research Center, Azerbaijan;
Conselho Nacional de Desenvolvimento Cient\'{\i}fico e Tecnol\'{o}gico (CNPq), Financiadora de Estudos e Projetos (Finep), Funda\c{c}\~{a}o de Amparo \`{a} Pesquisa do Estado de S\~{a}o Paulo (FAPESP) and Universidade Federal do Rio Grande do Sul (UFRGS), Brazil;
Ministry of Education of China (MOEC) , Ministry of Science \& Technology of China (MSTC) and National Natural Science Foundation of China (NSFC), China;
Ministry of Science and Education and Croatian Science Foundation, Croatia;
Centro de Aplicaciones Tecnol\'{o}gicas y Desarrollo Nuclear (CEADEN), Cubaenerg\'{\i}a, Cuba;
Ministry of Education, Youth and Sports of the Czech Republic, Czech Republic;
The Danish Council for Independent Research | Natural Sciences, the VILLUM FONDEN and Danish National Research Foundation (DNRF), Denmark;
Helsinki Institute of Physics (HIP), Finland;
Commissariat \`{a} l'Energie Atomique (CEA) and Institut National de Physique Nucl\'{e}aire et de Physique des Particules (IN2P3) and Centre National de la Recherche Scientifique (CNRS), France;
Bundesministerium f\"{u}r Bildung und Forschung (BMBF) and GSI Helmholtzzentrum f\"{u}r Schwerionenforschung GmbH, Germany;
General Secretariat for Research and Technology, Ministry of Education, Research and Religions, Greece;
National Research, Development and Innovation Office, Hungary;
Department of Atomic Energy Government of India (DAE), Department of Science and Technology, Government of India (DST), University Grants Commission, Government of India (UGC) and Council of Scientific and Industrial Research (CSIR), India;
Indonesian Institute of Science, Indonesia;
Istituto Nazionale di Fisica Nucleare (INFN), Italy;
Japanese Ministry of Education, Culture, Sports, Science and Technology (MEXT), Japan Society for the Promotion of Science (JSPS) KAKENHI and Japanese Ministry of Education, Culture, Sports, Science and Technology (MEXT)of Applied Science (IIST), Japan;
Consejo Nacional de Ciencia (CONACYT) y Tecnolog\'{i}a, through Fondo de Cooperaci\'{o}n Internacional en Ciencia y Tecnolog\'{i}a (FONCICYT) and Direcci\'{o}n General de Asuntos del Personal Academico (DGAPA), Mexico;
Nederlandse Organisatie voor Wetenschappelijk Onderzoek (NWO), Netherlands;
The Research Council of Norway, Norway;
Commission on Science and Technology for Sustainable Development in the South (COMSATS), Pakistan;
Pontificia Universidad Cat\'{o}lica del Per\'{u}, Peru;
Ministry of Education and Science, National Science Centre and WUT ID-UB, Poland;
Korea Institute of Science and Technology Information and National Research Foundation of Korea (NRF), Republic of Korea;
Ministry of Education and Scientific Research, Institute of Atomic Physics, Ministry of Research and Innovation and Institute of Atomic Physics and University Politehnica of Bucharest, Romania;
Joint Institute for Nuclear Research (JINR), Ministry of Education and Science of the Russian Federation, National Research Centre Kurchatov Institute, Russian Science Foundation and Russian Foundation for Basic Research, Russia;
Ministry of Education, Science, Research and Sport of the Slovak Republic, Slovakia;
National Research Foundation of South Africa, South Africa;
Swedish Research Council (VR) and Knut \& Alice Wallenberg Foundation (KAW), Sweden;
European Organization for Nuclear Research, Switzerland;
Suranaree University of Technology (SUT), National Science and Technology Development Agency (NSDTA) and Office of the Higher Education Commission under NRU project of Thailand, Thailand;
Turkish Energy, Nuclear and Mineral Research Agency (TENMAK), Turkey;
National Academy of  Sciences of Ukraine, Ukraine;
Science and Technology Facilities Council (STFC), United Kingdom;
National Science Foundation of the United States of America (NSF) and United States Department of Energy, Office of Nuclear Physics (DOE NP), United States of America.    
\end{acknowledgement}

\bibliographystyle{utphys}   
\bibliography{biblio}

\newpage
\appendix
\section{The ALICE Collaboration}
\label{app:collab}

\begin{flushleft}

\bigskip 

S.~Acharya$^{\rm 142}$, 
D.~Adamov\'{a}$^{\rm 96}$, 
A.~Adler$^{\rm 74}$, 
J.~Adolfsson$^{\rm 81}$, 
G.~Aglieri Rinella$^{\rm 34}$, 
M.~Agnello$^{\rm 30}$, 
N.~Agrawal$^{\rm 54}$, 
Z.~Ahammed$^{\rm 142}$, 
S.~Ahmad$^{\rm 16}$, 
S.U.~Ahn$^{\rm 76}$
% ...

\bigskip

\bigskip 

\textbf{\Large Affiliation Notes}

\bigskip 

% ...
$^{\rm VII}$ Also at: University of Kansas, Lawrence, Kansas, United States\\

\bigskip

\bigskip 

\textbf{\Large Collaboration Institutes}

\bigskip 

% ...
$^{16}$ Department of Physics, Aligarh Muslim University, Aligarh, India\\
% ...
$^{30}$ Dipartimento DISAT del Politecnico and Sezione INFN, Turin, Italy\\
% ...
$^{34}$ European Organization for Nuclear Research (CERN), Geneva, Switzerland\\
% ...
$^{54}$ INFN, Sezione di Bologna, Bologna, Italy\\
% ...
$^{74}$ Johann-Wolfgang-Goethe Universit\"{a}t Frankfurt Institut f\"{u}r Informatik, Fachbereich Informatik und Mathematik, Frankfurt, Germany\\
% ...
$^{76}$ Korea Institute of Science and Technology Information, Daejeon, Republic of Korea\\
% ...
$^{81}$ Lund University Department of Physics, Division of Particle Physics, Lund, Sweden\\
% ...
$^{96}$ Nuclear Physics Institute of the Czech Academy of Sciences, \v{R}e\v{z} u Prahy, Czech Republic\\
% ...
$^{142}$ Variable Energy Cyclotron Centre, Homi Bhabha National Institute, Kolkata, India\\
% ...

\bigskip 

\end{flushleft} 
\end{document}